\documentclass[twocolumn,showpacs,preprintnumbers,aps,prx]{revtex4-1}

\usepackage{hyperref}
\hypersetup{
    backref =       true,
    pagebackref  =  true,
    colorlinks =    true,
    linkcolor =     [rgb]{0.0,0.0,0.8},
    anchorcolor =   [rgb]{0.0,0.0,0.8},
    citecolor =     [rgb]{0.0,0.0,0.8},
    filecolor =     [rgb]{0.0,0.0,0.8},
    urlcolor =      [rgb]{0.0,0.0,0.8},
    pdftitle=       {Title},
    pdfsubject=     {Title},
    pdfauthor=      {A. Uthor}
}

\usepackage[utf8]{inputenc}
\usepackage{amsmath}
\usepackage{graphicx}
 
\usepackage{txfonts}
\usepackage{bm}
\usepackage{natbib}
\usepackage{siunitx}
\usepackage{microtype}
\makeatother
\begin{document}

\title{Photothermally Induced Transparency}

\author{Jinyong Ma, Jiayi Qin, Geoff T.\ Campbell, Ruvi Lecamwasam, Kabilan Sripathy, Joe Hope, Ben C.\ Buchler, and Ping Koy Lam$^{1}$}\email{Ping.Lam@anu.edu.au}

\affiliation{Centre for Quantum Computation and Communication Technology, Department of Quantum Science, Research School of Physics and Engineering, The Australian National University, Canberra ACT 2601, Australia}

\begin{abstract}
Induced transparency is a common but remarkable effect in optics. It occurs when a strong driving field is used to render an otherwise opaque material transparent. The effect is known as electromagnetically induced transparency in atomic media and optomechanically induced transparency in systems that consist of coupled optical and mechanical resonators. In this work, we introduce the concept of photothermally induced transparency (PTIT). It happens when an optical resonator exhibits non-linear behavior due to optical heating of the resonator or its mirrors. Similar to the established mechanisms for induced transparency, PTIT can suppress the coupling between an optical resonator and a traveling optical field. We further show that the dispersion of the resonator can be modified to exhibit slow or fast light. Because of the relatively slow thermal response, we observe the bandwidth of the PTIT to be $2\pi\times15.9\,\rm{Hz}$ which theoretically suggests a group velocity of as low as \SI{5}{\meter/\second}.
\end{abstract}

\maketitle

\section{Introduction}
Electromagnetically induced transparency (EIT) can occur when a medium that would otherwise absorb a probe field is rendered transparent by altering the atomic state with a control field \cite{BollerObservationelectromagneticallyinduced1991, KasapiElectromagneticallyInducedTransparency1995, fleischhauer_electromagnetically_2005}. The control field creates two dressed states that destructively interfere, resulting in a transparency window for a resonant probe field. The first demonstration of EIT was performed by Boller et al. in 1991 \cite{BollerObservationelectromagneticallyinduced1991}, showing transmittance transparency in an atomic transition. This phenomenon was widely recognized and applied in the manipulation of photons \cite{LukinColloquiumTrappingmanipulating2003, fleischhauer_electromagnetically_2005}. The EIT technique was shown to reduce the group velocity of light dramatically in an ultracold atomic gas ($\SI{17}{\meter/\second}$ at \SI{589}{\nano\meter}) \cite{HauLightspeedreduction1999}, thereby trapping light pulses for a controlled period of time (up to $\sim 0.5$ ms)
\cite{PhillipsStorageLightAtomic2001}. Observations of this ultraslow light pulse were then reported in other solid \cite{TurukhinObservationUltraslowStored2001,BigelowSuperluminalSlowLight2003} and gaseous~\cite{KashUltraslowGroupVelocity1999, BajcsyStationarypulseslight2003,PeyronelQuantumnonlinearoptics2012} media.

An analogous phenomenon exists in optomechanical systems composed of coupled optical and mechanical resonators, known as optomechanically induced transparency  (OMIT)~\cite{weis_OMIT_opto, safavi-naeini_slow-light_opto, ma_OMIT-mechanical-drive_opto}. Beating between control and probe fields produces a radiation pressure force that oscillates at the resonant frequency of the mechanical resonator and induces coherent oscillation of the resonator. The motion of the mechanical resonator then modulates the control field and generates a sideband that destructively interferes with the probe field to produce transparency. In 2010, Weis et al. \cite{weis_OMIT_opto} presented a form of OMIT in a toroidal microcavity. By changing input control light power, they achieved a tunable OMIT transparency window from 50 to 500 kHz compared with the total cavity loss rate of 15 MHz. The OMIT effect was also observed in other optomechanical systems \cite{LiuElectromagneticallyInducedTransparency2013, KaruzaOptomechanicallyinducedtransparency2013, QinLinearnegativedispersion2015}, and its nonlinear version has been investigated\cite{xiong_higher-order_2012, kronwald_optomechanically_2013}. It has found potential applications in slow light \cite{safavi-naeini_slow-light_opto, ChangSlowingstoppinglight2011}, Kerr nonlinearities \cite{lu_quantum-criticality-induced_2013} and precision measurement \cite{zhang_charge-measure_opto}. The presence of optomechanically induced absorption or a narrow gain feature can additionally lead to causality-preserving superluminal propagation (group advance) \cite{BigelowSuperluminalSlowLight2003, BigelowObservationUltraslowLight2003, BrunnerDirectMeasurementSuperluminal2004}.

Here, we theoretically propose and experimentally demonstrate a transparency phenomenon induced by the photothermal effects in an optical cavity. In a similar manner to radiation pressure \cite{kippenberg_analysis_2005, safavi-naeini_slow-light_opto,guccione_scattering-free_2013}, the photothermal effects couples cavity optical path length to the intracavity power. This is due to the absorption of photons by the cavity mirrors leading to thermal expansion and refractive index change of the mirror coating and substrate. These photothermal effects can either decrease or increase the optical path length of the cavity depending on the interaction~\cite{konthasinghe_self-sustained_2017, AnOpticalbistabilityinduced1997}. Just as with radiation pressure, the modulation in cavity length caused by the photothermal effects gives rise to feedback between intracavity power and cavity length.

\begin{figure*}[htb] \centering
\includegraphics[width=0.95\textwidth]{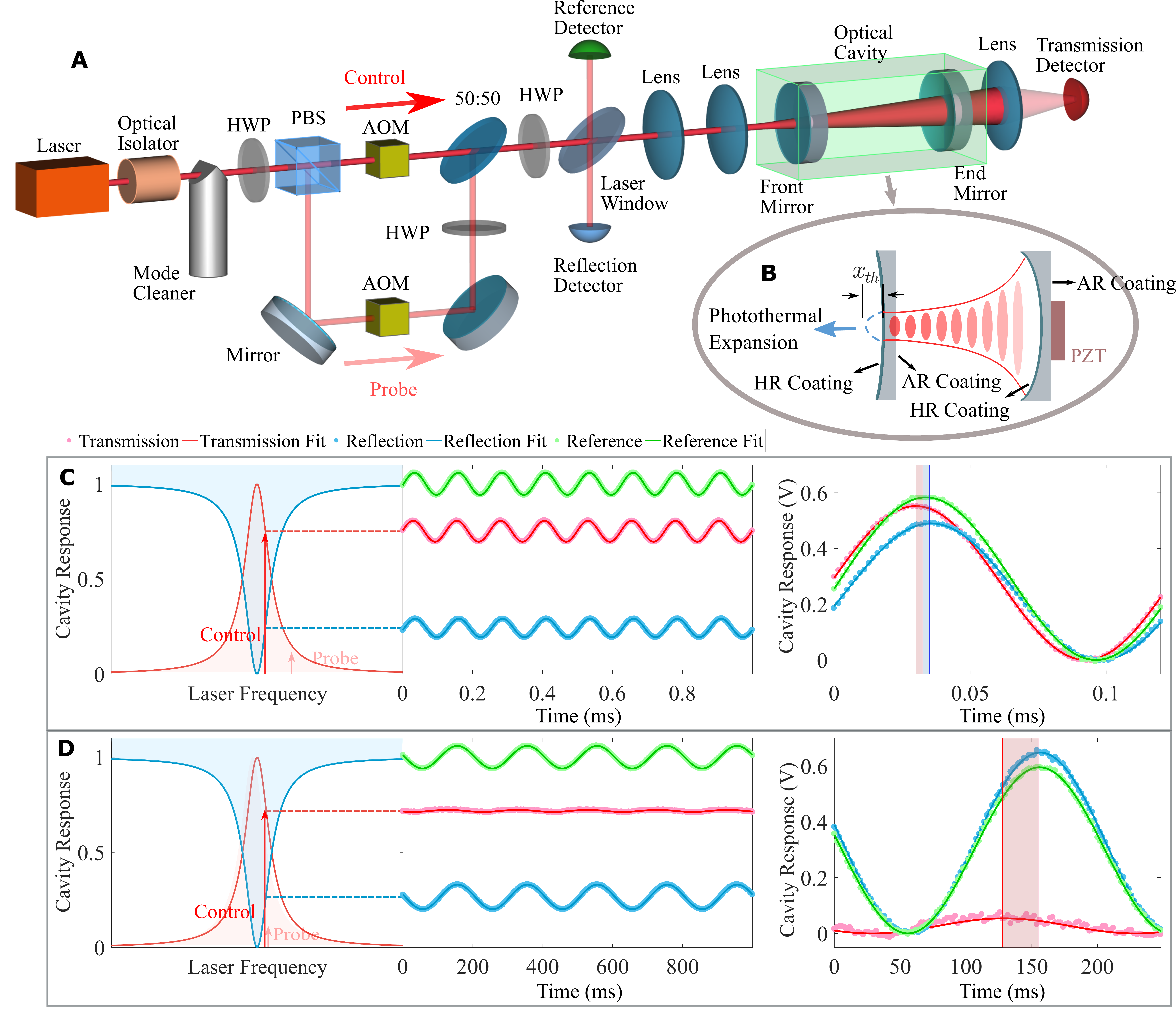} \caption{Experimental set-up and detection configuration. (\textbf{A}) Experimental set-up. A laser (\SI{1064}{\nano\meter}) is split into a strong laser (control) and a weak laser (probe) with their frequencies being modulated by two acousto-optic modulators (AOM) respectively. The two lasers are recombined by a 50:50 split and then coupled to a linear optical cavity. The substrate  (made from fused silica) of the cavity front mirror is placed inside the cavity such that the photothermal effects are enhanced. The polarization of the laser is tuned by a half-wave plate (HWP) to avoid birefringence effects. A laser window is used to pick up the cavity reflection (blue detector) and the reference signal (green detector). The transmitted power of the cavity is also detected by a photo-diode (red). (\textbf{B}) The strong laser is partially absorbed by the cavity mirrors, leading to the expansion of the mirror surfaces and the refractive index change of the substrate. The front mirror is orientated such that its high-reflectivity (HR) coating faces outward, and its anti-reflectivity (AR) coating faces inward. The end mirror attaches a piezoelectric actuator (PZT) used to scan the cavity length. (\textbf{\rm c})-(\textbf{D}) Reference (green), reflection (blue) and transmission (red) signals detected by the three optical detectors. The left panels present the spectral configuration of control and probe lasers. The middle panels show the detector signals normalized to cavity resonance. Each data set (dots) are fitted using a sine wave (solid lines). The curves on the right panels are shifted for clarity, which allows us to see the phase shift relative to the reference signal. The probe frequency relative to control frequency is set as $\Omega_p= 2\pi\times8\,\rm{kHz}$ and $\Omega_p = 2\pi\times 5\,\rm{Hz}$ for figures C and D respectively. The signals are measured at a control power of about \SI{90}{\milli\watt} and effective control detuning of about $2\pi\times 140\,\rm{kHz}$.}
\label{fig:1}
\end{figure*}

We investigate the dynamics of an optical cavity that is driven by both a bright control field and a weaker probe field. If the cavity exhibits photothermal effects, then interference between the two fields will lead to a modulation of the cavity optical path length at the frequency difference between the two driving fields. This length modulation will, in turn, generate Stokes and anti-Stokes optical sidebands. In the total cavity transmission spectrum including two-sidebands contribution, we find a near-unity dip with its efficiency and bandwidth being tunable via the power and effective detuning of the control field. We call the effect photothermally induced transparency (PTIT) for consistency with EIT and OMIT. In all three cases, the presence of a strong optical control field suppresses the coupling between a weaker optical probe and a resonance feature. For EIT, this produces increased transmission through an atomic media. For OMIT with a single-ended cavity, the effect is observed in the increased transmission \cite{weis_OMIT_opto} or decreased reflection \cite{safavi-naeini_slow-light_opto} from the coupling interface. Here, we consider PTIT in a two-ended cavity and the effect is observed as a \emph{decrease} in transmission. The term ``transparency'' is therefore counter-intuitive in this context but is consistent with the established terminology of OMIT.

In addition, we find that the transmission spectrum for the probe field is strongly modified in the vicinity of the control field frequency, showing a unique feature that cannot be found in EIT or OMIT. The spectrum exhibits both decreased and increased transmission, with a depth and width that are mediated by the control field intensity and the properties of the photothermal effects. This is the result of the destructive (or constructive) interference between the intracavity probe field and the scattered anti-Stokes (or Stokes) field. The spectrally narrow feature has a large dispersion that leads to an optically tunable delay and advancement of group velocity on the order of milliseconds. We note that, unlike EIT and OMIT, PTIT does not involve the interference of two quantum paths due to the dissipative nature of photothermal effects.

\section{Results}
\subsection{Bistability}
Our experimental setup is shown in Fig.~1A. A laser is split into a very intense control laser and a weak probe laser via a polarizing beam splitter (PBS). The detuning of the two beams is controlled using two acousto-optic modulators (AOMs). The two beams are then recombined and are injected into an optical cavity consisting of a convex front mirror and a concave end mirror. To measure the cavity response, we detect the transmitted and reflected light (denoted in red and blue, respectively), as well as a reference for the input intensity (denoted in green). A laser window is used as a beam pick-off for the reference beam and cavity reflection. Figures 1C-D show a sample of the collected data where the control is tuned near the cavity resonance. The beat note between the control and probe is detected and used to infer the amplitude of the probe and the relative phase between the two beams on transmission and reflection. When the frequency of the probe field is near that of the control field, the beat note is suppressed and phase-shifted in the transmitted signal.

There are several models that give quantitative descriptions of the photothermal effects in an optical cavity \cite{marino_canard_2006-2, AbdiQuantumoptomechanicsmultimode2012a}. Here, we use an empirical equation that has been demonstrated experimentally \cite{konthasinghe_self-sustained_2017}. We consider a cavity mode ($a$) which is driven by a strong control field with frequency $\omega_{\rm con}$ and power $P_{\rm con}$. The cavity mode is also driven by a much weaker probe field with frequency $\omega_{\rm p}$ and power $P_{\rm p}$. We investigate the dynamics of the system using the equations of motion in the rotating frame of $\omega_{\rm con}$
\begin{eqnarray}
\dot{x}_{\rm th} & = & -\gamma_{\rm th}(x_{\rm th}+\beta P_{\rm c}), \label{eq:PTE}\\
\dot{a} & = & -[\kappa/2-i(\Delta+Gx_{\rm th})]a+\varepsilon_{\rm con}+\varepsilon_{\rm p}e^{-i\Omega_{\rm p}t}, \label{eq:aTE}
\end{eqnarray}
where $x_{\rm th}$ is the total cavity length change due to photothermal effects (including photothermal expansion and photothermal refractive index change), $\gamma_{\rm th}$ is the effective photothermal relaxation rate, and $\beta=\left\vert dx_{\rm th}/dP_{\rm c}\right\vert$ is the effective photothermal coefficient. The sign of $\beta$ here is negative due to the outwards expansion of the front cavity mirror and the refractive index increase of its substrate. The control field is detuned from the cavity resonance by $\Delta=\omega_{\rm con}-\omega_{\rm cav}$. The amplitude of the control field is given as $\varepsilon_{\rm con}=\sqrt{P_{\rm con}\kappa_{f}/\hbar\omega_{\rm cav}}$ where $\kappa_f$ is the loss of the front mirror. The frequency of the probe (amplitude $\varepsilon_{\rm p}=\sqrt{P_{\rm p}\kappa_{f}/\hbar\omega_{\rm cav}}$) is $\Omega_{\rm p} = \omega_{\rm p} - \omega_{\rm con}$ in the rotating frame of the control frequency. The total loss rate of the cavity, $\kappa$, includes an external loss rate and an intrinsic loss rate. The intracavity power is $P_{\rm c}=\hbar\omega_{\rm cav}\left\vert a\right\vert ^{2}/\tau_{\rm cav}$, where $\tau_{\rm cav}=2L_{\rm c}/c$ is the cavity round-trip time and $L_{\rm c}$ is the cavity length. The cavity mode and the photothermal effects are coupled at the rate $G=\omega_{\rm cav}/L_{\rm c}$.

In the case that the probe field is much weaker than the control field, we can linearize Eqs.~(\ref{eq:PTE})-(\ref{eq:aTE}) using the assumptions, $x_{\rm th}=x_{0}+\delta x_{\rm th}$ , $a=a_{0}+\delta a$, and $a^*=a_{0}^*+\delta a^*$. We obtain following  steady state solutions after doing the linearization
\begin{eqnarray}
x_{0} & = & -\alpha\left\vert a_{0}\right\vert ^{2}, \label{eq:steadyx}\\
a_{0} & = & \frac{\varepsilon_{\rm con}}{\kappa/2-i(\Delta+Gx_{0})}, \label{eq:steadya}
\end{eqnarray}
and the linearized dynamical equations:
\begin{eqnarray}
\dot{\delta x_{\rm th}} & = & -\gamma_{\rm th}[\delta x_{\rm th}+\alpha(a_{0}\delta a^{*}+a_{0}^{*}\delta a)], \label{eq:dif1}\\
\delta\dot{a} & = & -\kappa\delta a/2+i\Delta_0\delta a+iGa_{0}\delta x_{\rm th}+\varepsilon_{\rm p}e^{-i\Omega_p t},  \label{eq:dif2}
\end{eqnarray}
where $\alpha=\beta\hbar\omega_{\rm cav}/\tau_{\rm cav}$, and $\Delta_0 = \Delta+Gx_0$ is the effective detuning of the control laser from the cavity resonance. We look at the steady-state solutions first before moving towards the analysis of the system dynamics.

\begin{figure}[htb] 
\includegraphics[width=0.48\textwidth]{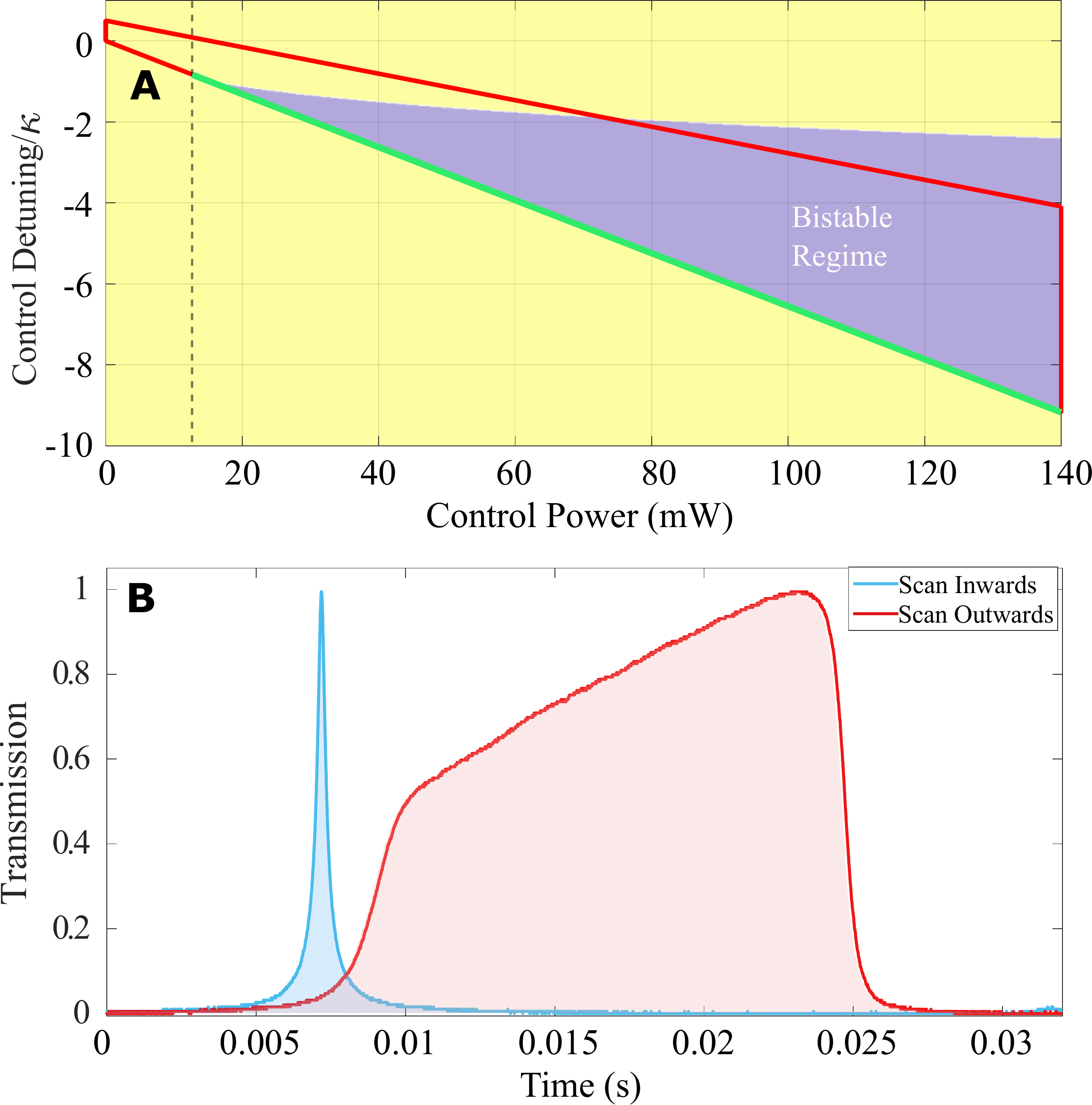} \caption{Stability of the system induced by the photothermal effects. (\textbf{A}) The stability map of the system, with the blue region being the bistability and yellow region being single stability. The parameter regime enclosed with solid curves corresponds to the one of Fig. 6. The green line represents the transition between single stability and bistability. (\textbf{B}) Experimental observation of optical bistability induced by the photothermal effects. At a control power of \SI{160}{\milli\watt}, the cavity response depends on the scanning direction. We observe the self-locking effect when moving the cavity mirror outwards and anti-self locking effect when moving the mirror inwards.}
\label{fig:2}
\end{figure}

The cavity resonance shift due to the photothermal effects is proportional to the cavity length change $x_0$ as indicated by Eq.~(\ref{eq:steadya}). Also, $x_0$ is linearly linked to the intracavity power as shown in Eq.~(\ref{eq:steadyx}). We can combine Eqs.~(\ref{eq:steadyx}) and (\ref{eq:steadya}) to obtain a cubic equation for $x_0$.
If the cubic equation has only one real root, then the system has only one steady state. If there are three distinct real roots, then the system is in a bistable state where two solutions are stable and the other one is not. Figure 2A maps stability against the free parameters of control field detuning and power, with the blue region representing the presence of the bistable state and the yellow region being the single-stability regime. Our following experiments run within the bistability regime where the cavity can be self-stabilized under a blue-detuned control without any external active feedback control.

\begin{figure*}[htb] \centering
\includegraphics[width=0.95\textwidth]{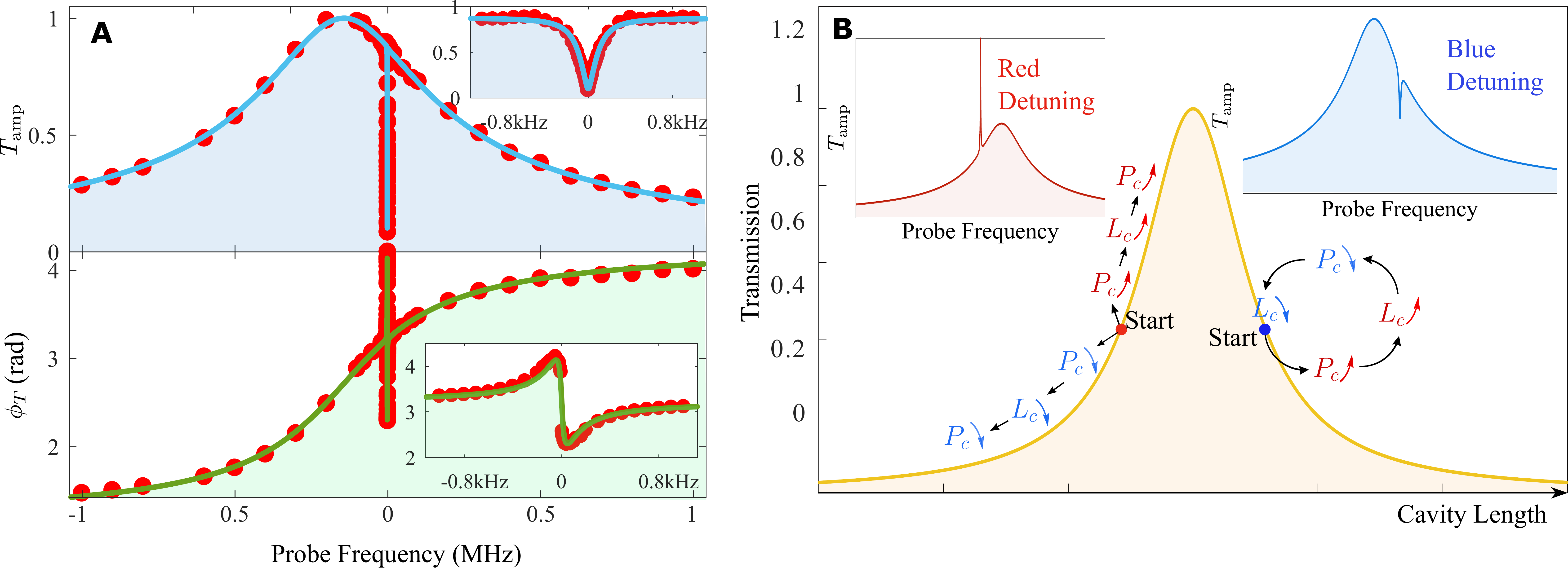} \caption{Observation of the photothermally induced transparency and an intuitive picture on self-locking. (\textbf{A}) Photothermally induced transparency observed in measured cavity transmission (dots) at a control power of about \SI{90}{\milli\watt}. We set $\Delta_0 = 0.28\kappa$ by manually tuning the piezoelectric actuator. Solid lines correspond to the model fits of experimental data. We observe a sharp transmission dip in the amplitude response $T_{\rm amp}$ of the dominant time-varying cavity transmission (top panels). The phase $\phi_T$ of the transmission is greatly altered (bottom panels), implying a strong dispersion behavior of the system.  (\textbf{B}) The diagram of the ``Beat-locking'' picture. The fluctuation of intracavity power induced by the probe field tends to converge in the blue-detuned regime and diverge in the red-detuned regime. The insets illustrate the diagrams of transmission dip and amplification peak present at blue and red detunings respectively. Note that the red-detuned cavity can be stable only in the single-stability regime. Given that our experiment run in the bistable regime, the relevant amplification process is not observed for the red-detuned case.}
\label{fig:3}
\end{figure*}

To explore the steady state of the cavity during the experiment, we slowly scan the cavity length using the piezoelectric actuator attached to the end cavity mirror. At a control power of $P_{\rm con}=\SI{160}{\milli\watt}$, we observe optical bistability in the transmitted signal of the cavity, as shown in Fig. 2B. The cavity resonance is shifted and the typical Lorentzian response of a cavity is deformed due to the photothermal non-linear interaction~\cite{AnOpticalbistabilityinduced1997, CarmonDynamicalthermalbehavior2004}. There are two distinct paths for the cavity behavior depending on the scanning direction. The cavity is self-locked when increasing the cavity length via the actuator and is anti-self-locked when scanning from the other direction.

\subsection{Photothermally induced transparency}

We now look at the dynamic behavior of the system. If we consider the ansatzs, $\delta x_{\rm th} = qe^{-i\Omega_{\rm p} t}+q^{*}e^{i\Omega_{\rm p} t}$, $\delta a = A^{-}e^{-i\Omega_{\rm p} t}+A^{+}e^{i\Omega_{\rm p} t}$,
$\delta a^{*} = (A^{-})^{*}e^{i\Omega_{\rm p} t}+(A^{+})^{*}e^{-i\Omega_{\rm p} t}$, and insert them into Eqs.~(\ref{eq:dif1}) and (\ref{eq:dif2}), we obtain the solution of the first order,
\begin{eqnarray}
A^{-} & = & \frac{1+if(\Omega_{\rm p})}{[-i(\Delta_{0}+\Omega_{\rm p})+\kappa/2]+2\Delta_{0}f(\Omega_{\rm p})}\varepsilon_{\rm p}
\end{eqnarray}
with
\begin{eqnarray}
f(\Omega_{\rm p}) & = & \frac{G\gamma_{\rm th}\alpha\left|a_{0}\right|^{2}}{[i(\Delta_{0}-\Omega_{\rm p})+\kappa/2](i\Omega_{\rm p}-\varepsilon).}
\end{eqnarray}
The transmitted field is given as follows,
\begin{eqnarray}
t_c & = & \kappa_{\rm e}a = \kappa_{\rm e}(a_{0}+A^{-}e^{-i\Omega_{\rm p}t}+A^{+}e^{i\Omega_{\rm p}t}), \label{eq:tc}
\end{eqnarray}
where $\kappa_{\rm e}$ denotes the loss of the cavity end mirror. In a bare optical cavity response, there is only one sideband as $A^+ = 0$ (see Supplementary Material). In the presence of photothermal effects, the beating of the control and probe fields induces a periodic oscillation of the effective cavity length $\delta x_{\rm th}$. The oscillation of $\delta x_{\rm th}$ gives rise to anti-Stokes and Stokes scattering from the control field. This process of photothermal back action generates two sidebands at $\Omega_{\rm p}$ (probe field) and $-\Omega_{\rm p}$ inside the cavity. The sidebands are not negligible as they are close to cavity resonance. The beat frequency of the control and probe fields determines the time scale of the process. Given that $a_0 \gg A^-$, the dominant time-varying signal of the cavity transmission $|t_c|^2$ is obtained by neglecting the higher-order terms:
\begin{eqnarray}
T &=& \left\vert \frac{\kappa a_{0}^{*}A^{-}+\kappa a_{0}(A^{+})^{*}}{2a_
{0}\varepsilon_{\rm p}}\right\vert\cos(\Omega_{\rm p} t + \phi_T), \label{eq:T}
\end{eqnarray}
where $\phi_T$ indicates the phase of $T$. Experimentally, the amplitude $T_{\rm amp}$ and phase $\phi_T$  of this signal are extracted from the data presented in the right panels of Figs. 1C-D: the amplitude of the sinusoidal signal refers to $T_{\rm amp}$ and the phase difference between the reference and transmission signals indicates $\phi_T$. Note that $T$ excludes the constant background of $t_c$. We will focus our discussion on transmission signal $T$ as this carries all the information about the intra-cavity field. 

\begin{figure*}[htb] \centering
\includegraphics[width=0.95\textwidth]{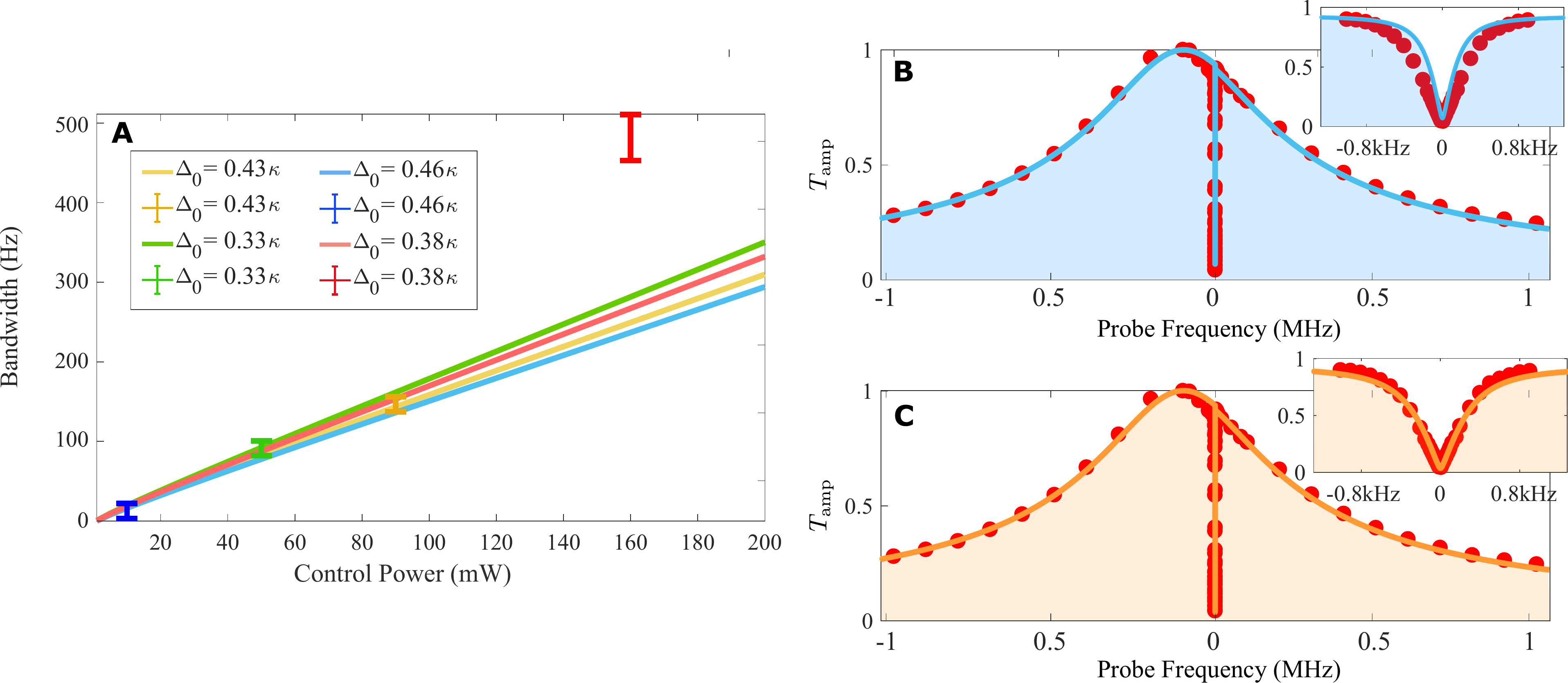} \caption{Theoretical and experimental results of the power dependence of the transparency bandwidth. (\textbf{A}) The bandwidth of transmission dip as a function of control laser power for several different control detunings. The bandwidth is obtained by fitting the dip to a Lorentzian function. Solid curves represent the model, and the points with error bars are the experimental results. Error bars indicate the 95 \% confidence interval. (\textbf{B})-(\textbf{\rm c}) Cavity transmission at high power of \SI{160}{\milli\watt} (red error bar). At high powers, the experimental result starts to deviate from the theory (panel B) as the increase of mirror temperature gives an increase of the photothermal coefficient. If taking into account the change of the photothermal coefficient, we can still fit the model to the data (panel C) and obtain a new photothermal coefficient.}
\label{fig:4}
\end{figure*}

The top panels of Fig.~3A present how the amplitude $T_{\rm amp}$ of the time-varying transmission depends on the probe frequency ($\Omega_{\rm p}/2\pi$) at two different control detunings. The red dots are the experimental data and the solid curves are the associated model fits of $T_{\rm amp}$. The broad resonance refers to the response of a bare cavity. A very narrow and near-unity dip is observed when the probe frequency is close to the control frequency (i.e., $\Omega_{\rm p} \approx 0$). The inset of the figure presents the details of the dip. The approximate profile of the transmission dip is a Lorentzian function (see Supplementary Material). The phase spectrum of the transmitted signal is shown in the bottom panel of Fig.~3A. A sharp change of the phase happens when the dip appears, i.e., $\Omega_{\rm p} \approx 0$. There is a good agreement between the model and experimental data, which allows us to precisely calibrate the photothermal parameters. We fit the model to data taken at several other control powers and detunings, which gives us fitting values of $\beta$ and $\gamma_{\rm th}$, i.e., $-1.8 \pm 0.2$ \SI{}{\pico\meter/\watt} and $2\pi\times(15.9\pm1.4)\,\rm{Hz}$ respectively. The error here is the standard deviation.

We also consider an intuitive picture based on self-locking to provide physical insights into the effects mentioned above (see Fig.~3B). We start the analysis from a single strong control field at $\Delta_0>0$. The cavity stays in a steady state under this single-frequency input. The blue point on the right side of the cavity resonance shown in Fig.~3B represents such a steady state in the case of an effective blue detuning. A secondary weak probe field, close to control frequency, then attempts to enter the cavity. The presence of the probe field can disrupt the stability of the cavity field due to the beat between the control and probe lasers. However, the following process prevents disruption from happening. When the control and probe laser are in phase, the presence of the probe field increases the overall intracavity power which in turn increases the cavity length (via photothermal effects). The increase of the cavity length then lowers the cavity power, which in turn cools the mirror and decreases the cavity length back towards what it was. As a result, the probe field fails to disrupt the cavity stability. This process gives rise to the transmission dip at the blue point (see Fig.~3B). We can do a similar analysis for the red dot located within the red-detuned regime, i.e., power up $\rightarrow$ cavity length increase $\rightarrow$ power up, and power down $\rightarrow$ cavity length decrease $\rightarrow$ power down. This process means that the probe can easily disturb the cavity stability and lead to amplification in intracavity power.

The bandwidth of the transmission dip is obtained by fitting it to a Lorentzian function. Figure~4A includes the theoretical and experimental results of the power dependence of the bandwidth at four different control detunings. The error bars indicate the standard deviation in the fit of the bandwidth. The theory shows that the bandwidth is linearly dependent on the control power. This is based upon the assumption that the photothermal coefficient $\beta$ does not change with the increase of the mirror temperature. The experimental data agree well with the theory at low control powers. There is, however, a disagreement at the control power of $\SI{160}{\milli\watt}$ (data of red dot) since the increased mirror temperature increases the value of $\beta$. The transmission dip linked to the red point is given in Fig.~4B. The inset of Fig.~4B presents a clear discrepancy between the data and the model under the photothermal parameters that we calibrate at low powers. The experimental result implies that the power dependence of the bandwidth is nonlinear at high powers. It is noted that the model will be still valid when taking into account the modifications of the photothermal parameters due to mirror heating. The orange line in Fig. 4C is a new fit of the model to data and we still see a good agreement. We obtain a new photothermal coefficient of \SI{-3.4}{\pico\meter/\watt} at this control power.

\subsection{Group delay and advance of probe field}

The previous section discussed the total transmission of the cavity. It is also of interest to explore the modification of the intracavity probe field in the presence of photothermal effects.  As discussed earlier, two optical sidebands at $\Omega_{\rm p}$ (same as probe field) and $-\Omega_{\rm p}$ are generated inside the cavity since the cavity power is coupled to the optical path length via photothermal effects. When focusing on the behavior of the probe field, we will only look at the sideband of the frequency which is the same as the probe. From Eq.~(\ref{eq:tc}), the normalized probe transmission is obtained as:
\begin{eqnarray}
t_p  &=&  \frac{\kappa A^-}{2\varepsilon_{\rm p}} 
 =  \frac{[1+if(\Omega_{\rm p})]\kappa/2}{[-i(\Delta_{0}+\Omega_{\rm p})+\kappa/2]+2\Delta_{0}f(\Omega_{\rm p})}, \label{Eq:Tp}
\end{eqnarray}

In the absence of the control field, i.e.,  $f(\Omega_{\rm p})=0$, Eq.~(\ref{Eq:Tp}) is reduced to a Lorentzian form which is the typical profile of a bare cavity response. Experimentally, the amplitude and phase of the probe transmission are calibrated from the measurement of the amplitude and phase of cavity transmission $T$ (see Supplementary Material).

\begin{figure}[htb] 
\includegraphics[width=0.48\textwidth]{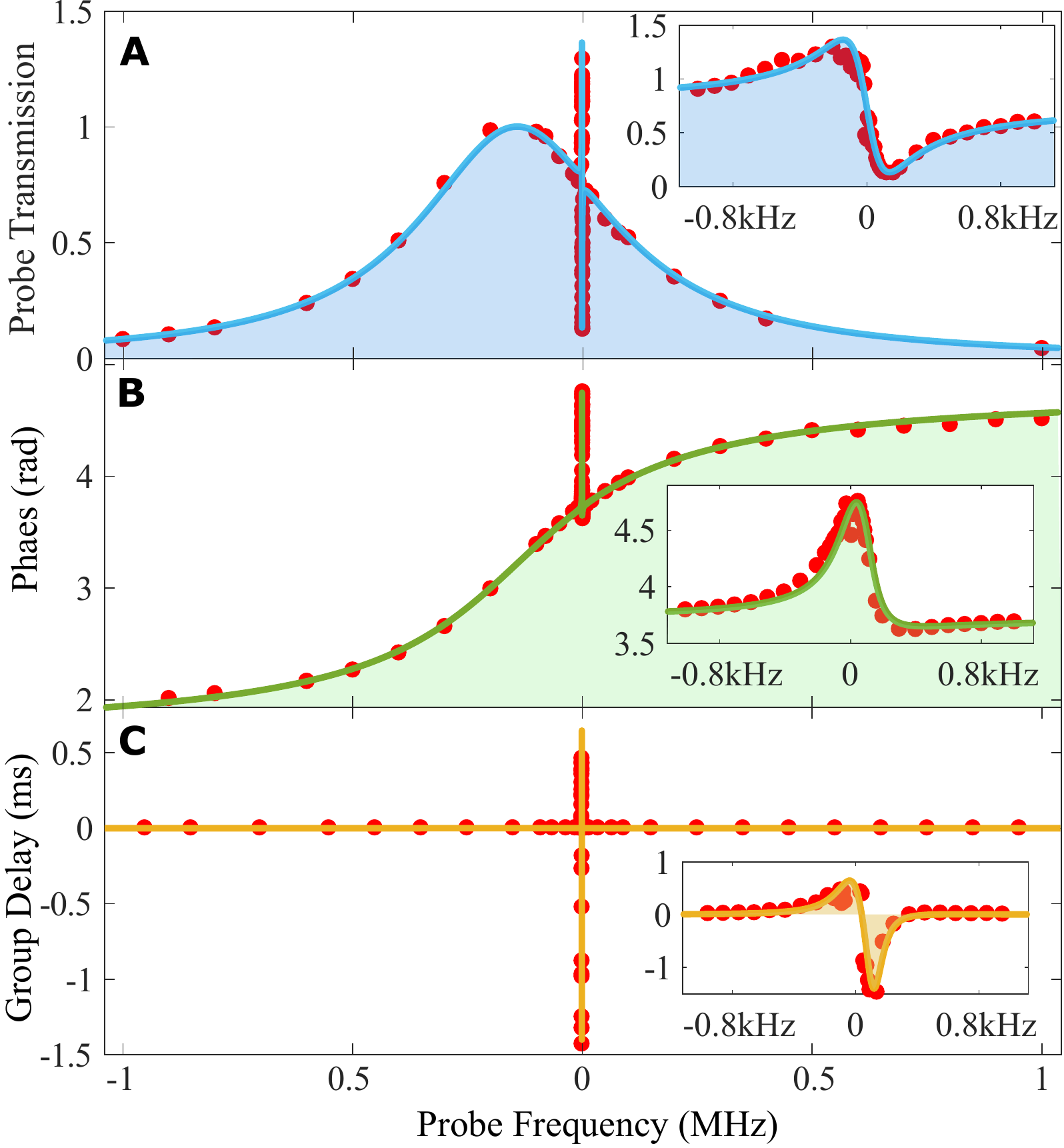} \caption{The probe transmission and group delay modified by a strong control field. (\textbf{A}) Probe transmission as a function of probe frequency ($\Omega_{\rm p}/2\pi$), including theoretical results (solid lines) and calibrated data (dots). Both a peak and a dip are present in a given control detuning. (\textbf{B}) The phase response of probe transmission. Alteration of phase at $\Omega_{\rm p}\approx0$ signifies strong cavity dispersion. (\textbf{\rm c}) Group delay of probe transmission. A positive value of delay implies a slow light effect while a negative one implies causality-preserving superluminal effect. $P_{\rm con} \approx \SI{90}{\milli\watt}$ and $\Delta_0 \approx 0.28\kappa$ are used for all panels.}
\label{fig:5}
\end{figure}

Figure~5A shows the theoretical prediction (solid curve) of the probe transmission and corresponding experimental result (red dots). The Lorentzian response of the transmitted probe field is modified in the presence of a strong control field. Under the same control detuning, the excitation of the intracavity probe field is either amplified or suppressed depending on the probe frequency. The inset of Fig.~5A shows the details of this effect. This behavior is similar to OMIT and EIT phenomena, though it is distinct. The transmission dip occurs at $\Omega_{\rm p} \approx 0$ where the probe frequency is very close to the control frequency while OMIT happens when the beat frequency of control and probe is equal to the resonant frequency of its mechanical resonator. In addition, the sharp signature in the probe transmission spectrum is asymmetric, with the simultaneous presence of a peak and a dip. In an optomechanical system, the transparency is present for a red-detuned control while the absorption appears in the blue-detuned regime.

We can use the scattering picture to explain this phenomenon. The beating of the control and probe fields induces the oscillation of the effective cavity length due to photothermal effects. The oscillation frequency is determined by the beating frequency as the ansatz $\delta x_{\rm th} = qe^{-i\Omega_{\rm p} t}+q^{*}e^{i\Omega_{\rm p} t}$ suggests. Furthermore, the amplitude and phase of the oscillation are controllable via the control and probe lasers. In turn, the oscillation leads to the Stokes- and anti-Stokes scattering of the control field. When $\Omega_{\rm p}>0$, the frequency of the probe field is the same as that of the scattered anti-Stokes field. Since the anti-Stokes field and the probe field are out of phase, their destructive interference suppresses the intracavity probe field and induces a transmission dip. When $\Omega_{\rm p}<0$, both the frequencies and the phases of the probe field and the Stokes field are the same. The interference of the two optical fields leads to amplification of the probe field.

The presence of the transmission dip or sharp absorption peak implies a strong modification of the cavity dispersion. The phase response of the transmitted probe field is shown in Fig.~5B.  A sharp change of the phase is observed at $\Omega_{\rm p} \approx0 $. The behavior of the probe phase gives a measure of the group delay or advance of the probe field as it travels through the cavity. We can obtain the group delay using the following two methods [see references \cite{weis_OMIT_opto} and \cite{safavi-naeini_slow-light_opto} respectively],
\begin{eqnarray}
\tau_{t}  =  \mathbf{R}\{\frac{-i}{t_p}\frac{dt_p}{d\Omega_{\rm p}}\}, \: or \:\:
\tau_{t} =  \frac{d\phi_{t_p}(\Omega_{\rm p})}{d\Omega_{\rm p}},
\end{eqnarray}
where $\phi_{t_p}(\Omega_{\rm p})$ is the phase of the probe transmission obtained from Eq.~(\ref{Eq:Tp}). The sign of $\tau_{t}$ determines the property of the light, that is, a positive and a negative sign imply slow light and fast light respectively \cite{BigelowSuperluminalSlowLight2003}. At control power of $P_{\rm con} \approx \SI{90}{\milli\watt}$ and effective control detuning of $\Delta_0 \approx 0.28\kappa$,  we observe a maximum group delay of about $\SI{0.6}{\milli\second}$ at $\Omega_{\rm p}<0$ and a maximum group advance of about $\SI{1.4}{\milli\second}$ at $\Omega_{\rm p}>0$ (see Fig.~5C). The simultaneous presence of the effects of slow and fast light is due to the asymmetric feature in the probe transmission spectrum, that is, it is a result of the photothermally induced transparency and absorption.

\begin{figure}[htb] 
\includegraphics[width=0.48\textwidth]{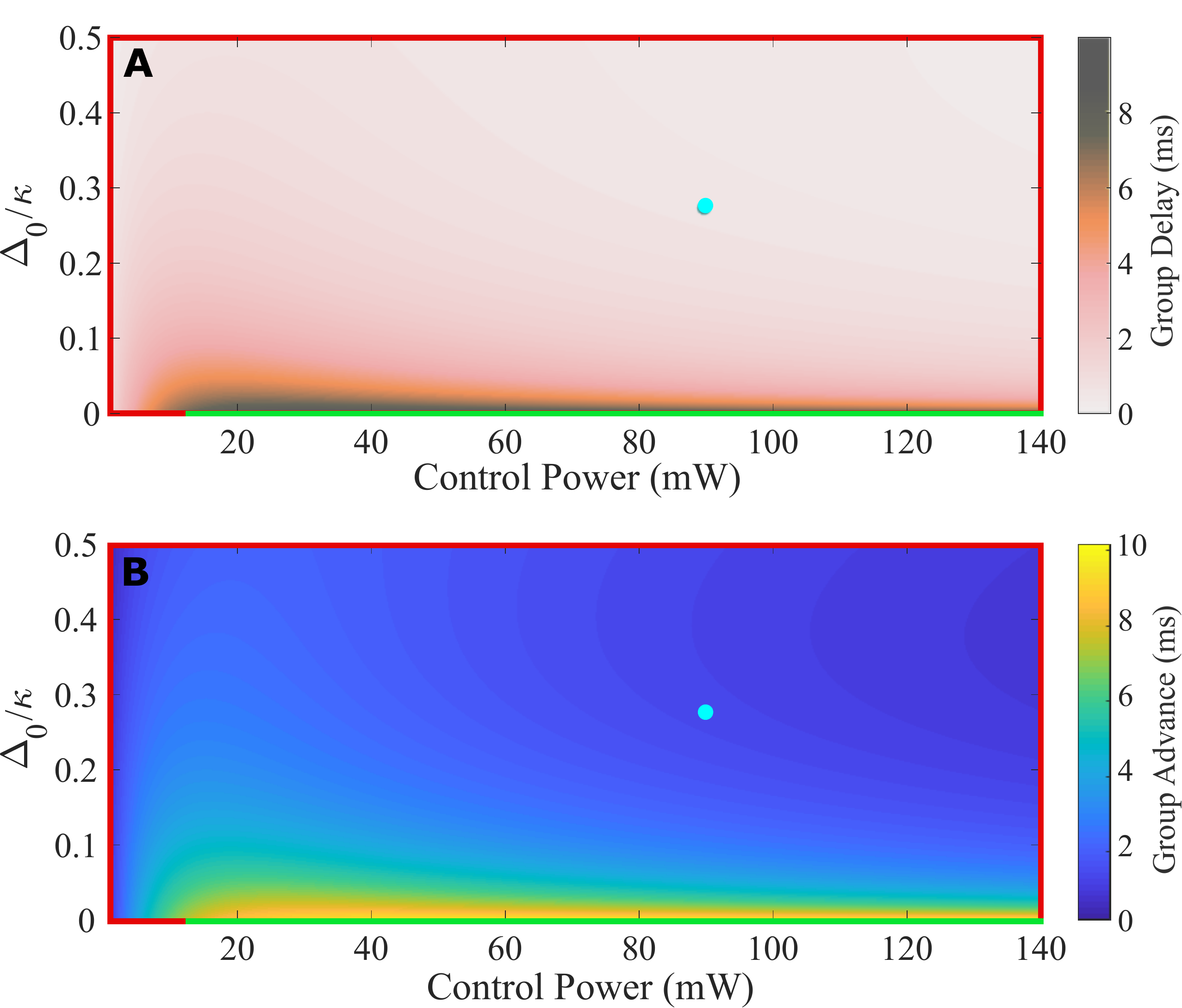} \caption{Theoretical predictions of group delay and advance. Theoretical prediction of group delay (\textbf{A}) and advance (\textbf{B}) as a function of control powers and effective control detunings ($\Delta_0$). The blue dots correspond to the experimental description in Fig.~5. Here we take the absolute value for group advance. Both delay and advance are tunable via control detuning and power. The green curves indicate the boundary between single stability and bistability, as mentioned in Fig. 2.}
\label{fig:6}
\end{figure}

The values of delay and advance are dynamically tunable via the intensity or detuning of the control laser. The theoretical prediction is shown in Fig.~6. The blue dots on the plots correspond to the case we discussed in Fig.~5. Here we assume that the photothermal coefficient remains constant as the mirror temperature increases. We can switch between slow and fast light effects easily by modulating the detuning of the control laser. The maximum delay can be about \SI{10}{\milli\second}, which is much longer than the ones achieved on OMIT \cite{safavi-naeini_slow-light_opto}. The cavity length $L_c$ is \SI{50}{\milli\meter}, which gives us a pulse propagation velocity of about \SI{5}{\meter/\second} (The group velocity is obtained from $v_g = L_c/\tau_t$ \cite{PoonSlowinglightFabryPerot2007, conti_nonlinear_2014}). The bandwidth of the pulse, however, is limited by the photothermal relaxation rate which in this case is about $2\pi\times15.9\,\rm{Hz}$. As mentioned earlier, we plot Fig.~6 in the parameter regime which is enclosed with solid curves in Fig.~2. The green curves suggest that huge delays and advances of the group velocity can occur at the transition from bistability to single stability. With regard to the superluminal effect, the peak of a narrow-band pulse is faster than light and travels through the cavity before it enters into the cavity. This case, however, occurs at the price of the distortion of the pulse. The signal (front of the pulse) is still sub-luminal and satisfies the principles of causality and relativity for the transfer of energy or information \cite{ThevenazSlowfastlight2008}.

\section{Discussion}
We are the first to propose and demonstrate the transparency phenomenon induced by photothermal effects. We apply a weak laser to probe the response of an optical cavity that is strongly driven by a control laser and exhibits photothermal effects. The total cavity transmission includes two sidebands as a result of the photothermal back action. We experimentally observe a narrow dip in the cavity transmission power spectrum which is in line with the theoretical prediction. The bandwidth of this dip is controllable via the control power and detuning. Furthermore, we report a near-unity dip as well as a sharp peak in the probe transmission. We also find a strong modification of the phase response of the probe field when the probe frequency is close to control frequency. Such an intense dispersion leads to a maximum group delay and a maximum group advance on the order of milliseconds. The delay and advance are capable of being dynamically tuned by control powers and effective control detunings. It is worth noting that the delay or advance of the group velocity is also determined by the photothermal parameters. Some materials have practically tunable photothermal parameters~\cite{catafesta_tunable_2006}. The photothermal relaxation rate can also be controlled by the size of the beam spot on the mirror. This can easily be achieved by adjusting the transverse electromagnetic mode via the change of the cavity length. Ultimately, this highly tunable group advance/delay phenomenon makes the photothermal effects attractive in the field of all-optical control. 

We would like to note that it might be difficult to extend the effects investigated here towards the quantum regime since the optical information transferred to the photothermal effects can dissipate into the environment. Unlike EIT and OMIT, the underlying mechanism for PTIT does not involve interference between two quantum paths. However, as with EIT and OMIT, PTIT arises due to destructive interference between a probe field and the anti-Stokes sideband of light scattered from a control field. This interference leads to a transparency window that is narrower than the cavity or absorption linewidth. Also, it has been experimentally demonstrated that the presence of photothermal effects can suppress the Brownian fluctuations of a microlever \cite{metzger_cavity_2004}. A recent theoretical work \cite{pinard_quantum_2008} proposed that the photothermal effects can cool a mechanical resonator down close to its quantum ground state in the bad-cavity limit. These works suggest that it may be possible to achieve a quantum version of photothermally induced transparency.

Considering that the photothermal-cavity interactions can either set a fundamental limit to metrology applications \cite{cerdonio_thermoelastic_2001, de_rosa_experimental_2002} or offer an effective way of suppressing Brownian noise \cite{metzger_cavity_2004, GiganSelfcoolingmicromirrorradiation2006}, the characterization of photothermal effects is crucial for cavity-based experiments requiring high sensitivity. A straightforward application of the PTIT effect is to characterize the photothermal parameters. One can easily set up an experiment similar to ours and fit the transmission data to our model to extract the photothermal parameters. In addition, all effects reported in our work are easy to access experimentally and thus show a  convenient way towards applications on classical signal processing, e.g., optical amplification and filtering.

\section{Materials and Methods}
The cavity mirrors of our system were attached to a hollow Invar cylinder to form a resonator that has reduced thermal variations and acoustic noise. The beam waist of the cavity field is close to the front mirror such that the laser intensity at the front mirror is higher than at the end mirror so that the photothermal effects on the front mirror is dominant. We oriented the front mirror with the high-reflectivity coating facing outwards so that the intra-cavity field passes through the substrate, as shown in Fig.~1B. In this configuration, the substrate of the front mirror was heated by the absorption of intracavity photons, leading to both a change in the refractive index of the substrate and outward thermal expansion of the mirror surface. This allows us to explore the photothermal effects at low laser powers. The substrate of our cavity mirrors was made from fused silica. One can use another material with a higher absorption coefficient as the substrate (e.g., BK7), which will further lower the power requirements for the experiment.

The system parameters are calibrated as follows: cavity length  $L_{\rm c} = \SI{0.05}{\meter}$, cavity finesse $\mathcal{F} = 5760$, cavity decay $\kappa= 2\pi\times530\,\rm{kHz}$, photothermal coefficient $\beta = \SI{-1.8}{\pico\meter/\watt}$, photothermal relaxation rate $\gamma_{\rm th} = 2\pi\times15.9\,\rm{Hz}$, and laser wavelength $\lambda_{\rm c} = \SI{1064}{\nano\meter}$.

\section*{Acknowledgments}
\noindent\textbf{Funding:} This research was funded by the Australian Research Council Centre of Excellence CE110001027 and the Australian Government Research Training Program Scholarship. PKL acknowledges support from the ARC Laureate Fellowship FL150100019.
\textbf{Author contributions:} J.M. and P.L. conceived the idea; J.M. and R.L. did the theoretical calculations. J.M. designed and fabricated the device. J.M., J.Q., and G.C. performed the experiment. J.M. wrote the manuscript and all authors contributed to the manuscript. P.L. supervised the work.
\textbf{Competing interests:} The authors declare that they have no competing interests. 
\textbf{Data and materials availability:} All data needed to evaluate the conclusions in the paper are present in the paper and/or the Supplementary Materials. Additional data related to this paper may be requested from the authors.

\section*{Supplementary materials}
\noindent Section S1. Modeling\\
Section S2. Shape of the transparency window of the cavity transmission\\
Section S3. Calibration of probe transmission\\
Section S4. A simplified solution

\onecolumngrid
\appendix
\clearpage
\section{Solution of linearized equation} \label{app:solu_PTIT}
We investigate the dynamics of our system using the equations of motion in the rotating frame of control frequency $\omega_{\rm con}$: 
\begin{eqnarray}
\dot{x}_{\rm th} & = & -\gamma_{\rm th}(x_{\rm th}+\beta P_{\rm c}), \label{eq:PTE_app}\\
\dot{a} & = & -[\kappa/2-i(\Delta+Gx_{\rm th})]a+\varepsilon_{\rm con}+\varepsilon_{p}e^{-i\Omega_{p}t}, \label{eq:aTE_app}
\end{eqnarray}
where $x_{\rm th}$ is the total cavity length change due to photothermal effects, $\gamma_{\rm th}$ is the photothermal relaxation rate, and $\beta=\left\vert dx_{\rm th}/dP_{\rm c}\right\vert$ is the photothermal expansion coefficient. The sign of $\beta$ here is negative due to the outwards expansion of the front cavity mirror and the refractive index change of the its substrate. The control field (amplitude $\varepsilon_{\rm con}=\sqrt{P_{\rm con}\kappa_{\rm f}/\hbar\omega_{\rm cav}}$) is detuned from the cavity resonance by $\Delta=\omega_{\rm con}-\omega_{\rm cav}$, and the frequency of the probe (amplitude $\varepsilon_{\rm p}=\sqrt{P_{\rm p}\kappa_{\rm f}/\hbar\omega_{\rm cav}}$) is $\Omega_{\rm p} = \omega_{\rm p} - \omega_{\rm con}$ in the rotating frame of the control frequency. The total loss rate of the cavity, $\kappa$, includes an external loss rate and an intrinsic loss rate. The intracavity power is $P_{\rm c}=\hbar\omega_{\rm cav}\left\vert a\right\vert ^{2}/\tau_{\rm cav}$, where $\tau_{\rm cav}=2L_{\rm c}/c$ is the cavity round-trip time and $L_{\rm c}$ is the cavity length. The cavity mode and the photothermal effects are coupled at the rate $G=\omega_{\rm cav}/L_{\rm c}$.

We can linearize Eqs.~(\ref{eq:PTE_app})-(\ref{eq:aTE_app}) using the assumptions that $x_{\rm th}=x_{0}+\delta x_{\rm th}$ , $a=a_{0}+\delta a$, and $a^*=a_{0}^*+\delta a^*$ in the case that the probe field is much weaker than the control field. We obtain following  steady state solutions after doing the linearization: 
\begin{eqnarray}
x_{0} & = & -\alpha\left\vert a_{0}\right\vert ^{2}, \label{eq:steadyx_appPTIT}\\
a_{0} & = & \frac{\varepsilon_{\rm con}}{\kappa/2-i(\Delta+Gx_{0})}, \label{eq:steadya_appPTIT}
\end{eqnarray}
where $\alpha=\beta\hbar\omega_{\rm cav}/\tau_{\rm cav}$, and the linearized dynamical equations:
\begin{eqnarray}
\dot{\delta x_{\rm th}} & = & -\gamma_{\rm th}[\delta x_{\rm th}+\alpha(a_{0}\delta a^{*}+a_{0}^{*}\delta a)], \label{eq:dif1_app}\\
\delta\dot{a} & = & -\kappa\delta a/2+i\Delta_0\delta a+iGa_{0}\delta x_{\rm th}+\varepsilon_{p}e^{-i\Omega_p t},  \label{eq:dif2_app}
\end{eqnarray}
where $\Delta_0 = \Delta+Gx_0$ is the effective detuning of the control laser from the cavity resonance.

Considering the following ansatzs,
\begin{eqnarray}
\delta x_{th} & = & qe^{-i\Omega_{\rm p} t}+q^{*}e^{i\Omega_{\rm p} t}\\
\delta a & = & A^{-}e^{-i\Omega_{\rm p} t}+A^{+}e^{i\Omega_{\rm p} t}\\
\delta a^{*} & = & (A^{-})^{*}e^{i\Omega_{\rm p} t}+(A^{+})^{*}e^{-i\Omega_{\rm p} t}
\end{eqnarray}
we obtain:
\begin{eqnarray}
(i\Omega_{\rm p}-\gamma_{\rm th})q & = & \gamma_{\rm th}\alpha[a_{0}(A^{+})^{*}+a_{0}^{*}(A^{-})] \label{q}\\
(-i(\Delta_{0}+\Omega_{\rm p})+\kappa/2)A^{-} & = & iGa_{0}q+\varepsilon_{\rm p} \label{AN}\\
(i(\Delta_{0}-\Omega_{\rm p})+\kappa/2)(A^{+})^{*} & = & -iGa_{0}^{*}q \label{AP}
\end{eqnarray}

We can easily get the solution as
\begin{eqnarray}
A^{-} & = & \frac{1+if(\Omega_{\rm p})}{[-i(\Delta_{0}+\Omega_{\rm p})+\kappa/2]+2\Delta_{0}f(\Omega_{\rm p})}\varepsilon_{\rm p} \label{A-}
\end{eqnarray}
with
\begin{eqnarray}
f(\Omega_{\rm p}) & = & \frac{G\gamma_{th}\alpha\left|a_{0}\right|^{2}}{[i(\Delta_{0}-\Omega_{\rm p})+\kappa/2](i\Omega_{\rm p}-\gamma_{\rm th})}
\end{eqnarray}

When there is no photothermal interaction, cavity length does not change, i.e., $q = 0$. From Eq. (\ref{AP}), we get $A^{+} = 0$ in this case, implying that there is only one sideband inside the cavity, which is the intracavity probe field. 

We thus have the cavity reflection as follows
\begin{eqnarray}
r & = & S_{in}-\kappa a/2 \nonumber\\
 & = & \varepsilon_{con}+\varepsilon_{p}e^{-i\Omega_{\rm p}t}-\frac{\kappa}{2}(a_{0}+A^{-}e^{-i\Omega_{\rm p}t}+A^{+}e^{i\Omega_{\rm p}t}) \nonumber\\
 & = & \varepsilon_{con}-\frac{\kappa}{2}a_{0}+(\varepsilon_{\rm p}-\frac{\kappa}{2}A^{-})e^{-i\Omega_{\rm p}t}-\frac{\kappa}{2}A^{+}e^{i\Omega_{\rm p}t}
\end{eqnarray}

\begin{eqnarray}
|r|^{2} 
 & = & (\varepsilon_{con}-\frac{\kappa}{2}a_{0})(\varepsilon_{con}-\frac{\kappa}{2}a_{0}^{*})+(\varepsilon_{\rm p}-\frac{\kappa}{2}A^{-})(\varepsilon_{\rm p}-\frac{\kappa}{2}(A^{-})^{*})+\frac{\kappa}{2}A^{+}(A^{+})^{*} \nonumber\\
 &  & +[(\varepsilon_{con}-\frac{\kappa}{2}a_{0}^{*})(\varepsilon_{\rm p}-\frac{\kappa}{2}A^{-})-(\varepsilon_{con}-\frac{\kappa}{2}a_{0})\frac{\kappa}{2}(A^{+})^{*}]e^{-i\Omega_{\rm p}t}\nonumber\\
 &  &+[(\varepsilon_{con}-\frac{\kappa}{2}a_{0})(\varepsilon_{\rm p}-\frac{\kappa}{2}A^{-})^{*}-(\varepsilon_{con}-\frac{\kappa}{2}a_{0}^{*})\frac{\kappa}{2}A^{+}]e^{i\Omega_{\rm p}t} \nonumber\\
 &  & -(\varepsilon_{\rm p}-\frac{\kappa}{2}A^{-})\frac{\kappa}{2}(A^{+})^{*}e^{-2i\Omega_{\rm p}t}-(\varepsilon_{\rm p}-\frac{\kappa}{2}A^{-})^{*}\frac{\kappa}{2}A^{+}e^{2i\Omega_{\rm p}t} 
\end{eqnarray}

The cavity transmission is also calculated as follows
\begin{eqnarray}
t & = & \kappa a/2 \nonumber = \frac{\kappa}{2}(a_{0}+A^{-}e^{-i\Omega_{\rm p}t}+A^{+}e^{i\Omega_{\rm p}t}) \label{tc}
\end{eqnarray}

\begin{eqnarray}
|t|^{2} & = & \frac{\kappa^{2}}{4}(a_{0}+A^{-}e^{-i\Omega_{\rm p}t}+A^{+}e^{i\Omega_{\rm p}t})[a_{0}^{*}+(A^{-})^{*}e^{i\Omega_{\rm p}t}+(A^{+})^{*}e^{-i\Omega_{\rm p}t}] \nonumber\\
 & = & \frac{\kappa^{2}}{4}[a_{0}a_{0}^{*}+A^{-}(A^{-})^{*}+A^{+}(A^{+})^{*} \nonumber\\
 &  & +(a_{0}^{*}A^{-}+a_{0}(A^{+})^{*})e^{-i\Omega_{\rm p}t}+(a_{0}(A^{-})^{*}+a_{0}^{*}A^{+})e^{i\Omega_{\rm p}t} \nonumber\\
 &  & +A^{-}(A^{+})^{*}e^{-2i\Omega_{\rm p}t}+A^{+}(A^{-})^{*}e^{2i\Omega_{\rm p}t}]
\end{eqnarray}

Here we assume that the losses of the front ($\kappa_{\rm f}$) and end ($\kappa_{\rm e}$) mirrors contribute equally to the total loss ($\kappa$). Given that $a_0 \gg A^-$, the normalized oscillation part (T) of the transmitted signal is easily obtained by ignoring the higher-order terms: 
\begin{eqnarray}
T &=& \left\vert \frac{\kappa a_{0}^{*}A^{-} +\kappa a_{0}(A^{+})^{*}}{2a_
{0}\varepsilon_{\rm p}}\right\vert\cos(\Omega_p t + \phi_T) \label{eq:T_app}
\end{eqnarray}
where $\phi_T$ indicates the phase behavior of the cavity transmission.

\section{Shape of the transparency window of the cavity transmission} \label{app:lor_PTIT}
\paragraph{}We can see that the transmission windows in Fig 3 a,b, and Fig 4b follow a Lorentzian shape. In this section, we will derive an expression for this. We begin with the amplitude of cavity transmission:
\begin{equation}
  T_{amp}=\left\lvert\frac{\kappa a_0^*A^-+\kappa a_0(A^+)^*}{2a_0\varepsilon_p}\right\rvert.
\end{equation}
The square of this may be expanded as
\begin{equation}\label{eq:TExpanded}
  \begin{aligned}
  |T_{amp}|^2&=\left[\kappa ^2 \left(\Omega_p ^2+\gamma_{th} ^2\right) \left(4 (\Delta_0-\Omega_p )^2+\kappa ^2\right)\right] / \\
  &\;\;\;\;\left[64 \alpha ^2 |a_0|^2 \Delta_0^2 G^2 \gamma_{th} ^2-16 \alpha  |a_0|^2 \Delta_0 G \gamma_{th}  \left(4 \Delta_0^2 \gamma_{th} +\kappa ^2 \gamma_{th} -4 \Omega_p ^2 (\kappa +\gamma_{th} )\right)\right. \\
  &\;\;\;\;\left.+\left(\Omega_p ^2+\gamma_{th} ^2\right) \left(4 (\Delta_0-\Omega_p )^2+\kappa ^2\right) \left(4 (\Delta_0+\Omega_p )^2+\kappa ^2\right)\right].
  \end{aligned}
\end{equation}
We wish to approximate $|T_{amp}^2|$ over the transmission window, which occurs when $\Omega_p$ is of the same order as $\gamma_{th}$. To analyse this regime we introduce dimensionless parameters:
\begin{equation}
    \tilde{\Delta}_0 = \frac{\Delta_0}{\kappa}; \; \tilde{\Omega}_p = \frac{\Omega_p}{\kappa}; \; \tilde{G} = \frac{G\alpha}{\kappa}; \; \kappa' = \frac{\kappa}{\gamma_{th}}.
\end{equation}
In terms of these we have
\begin{equation}
  \begin{aligned}
    |T_{amp}|^2 &= \left[(1+4(\tilde{\Delta}_0-\tilde{\Omega}_p)^2)(1+(\kappa'\tilde{\Omega}_p)^2)\right] / \\
      &\;\;\;\;\left[64|a_0|^2\tilde{G}^2\tilde{\Delta}_0^2+16|a_0|^2\tilde{G}\tilde{\Delta}_0(-1-4\tilde{\Delta}_0^2+4(1+\kappa')\tilde{\Omega}_p^2)\right. \\
      &\left.\;\;\;\;+(1+4(\tilde{\Delta}_0-\tilde{\Omega}_p)^2)(1+(\kappa'\tilde{\Omega}_p)^2)(1+4(\tilde{\Delta}_0+\tilde{\Omega}_p)^2)\right].
  \end{aligned}
\end{equation}
We now approximate $|\tilde{\Omega}_p|\ll|\tilde{\Delta}_0|$, as our detuning $\Delta_0$ is of order $\kappa$ while $\Omega_p$ is of order $\gamma_{th}$. This corresponds to sending 
\begin{equation}
  \begin{aligned}
    \tilde{\Delta}_0\pm\tilde{\Omega}_p &\rightarrow \tilde{\Delta}_0, \\
    -4\tilde{\Delta}_0^2+4(1+\kappa')\tilde{\Omega}_p^2&\rightarrow -4\tilde{\Delta}_0^2,
  \end{aligned}
\end{equation}
in which case the transmission window becomes (returning to the original variables)
\begin{equation}\label{eq:TransmissionWindow}
  T_{window}=\frac{\kappa^2(4\Delta_0^2+\kappa^2)(\gamma_{th}^2+\Omega_p^2)}{\gamma_{th}^2\left(4\Delta_0(\Delta_0-2|a_0|^2G\alpha)+\kappa^2\right)^2+(4\Delta_0^2+\kappa^2)^2\Omega_p^2}.
\end{equation}
We can recover the Lorentzian shape by subtracting (\ref{eq:TransmissionWindow}) from the response of a bare cavity evaluated at $\Omega_p=0$:
\begin{equation}
  \frac{1}{1+\left(\frac{\Delta_0}{\kappa/2}\right)^2}-\frac{\kappa^2(4\Delta_0^2+\kappa^2)(\gamma_{th}^2+\Omega_p^2)}{\gamma_{th}^2\left(4\Delta_0(\Delta_0-2|a_0|^2G\alpha)+\kappa^2\right)^2+(4\Delta_0^2+\kappa^2)^2\Omega_p^2}
\end{equation}
\begin{equation}
  =\frac{16 \alpha  |a_0|^2 \Delta_0 G \kappa ^2 \gamma_{th} ^2 \left(4 \alpha  |a_0|^2 \Delta_0 G-4 \Delta_0^2-\kappa ^2\right)}{\gamma_{th} ^2 \left(4 \Delta_0^2+\kappa ^2\right) \left(4 \Delta_0 (\Delta_0-2 \alpha  |a_0|^2G)+\kappa ^2\right)^2+\omega ^2 \left(4 \Delta_0^2+\kappa ^2\right)^3}
\end{equation}

\section{Calibration of probe transmission} \label{app:probe_PTIT}
If focusing on the behavior of the probe field, we will only look at the sideband of the frequency which is the same as the probe. From Eq. (\ref{tc}), the normalized probe transmission is obtained as:
\begin{eqnarray}
t_p & = & \frac{\kappa A^-}{2\varepsilon_{\rm p}} \nonumber\\
& = & \frac{[1+if(\Omega_p)]\kappa/2}{[-i(\Delta_{0}+\Omega_p)+\kappa/2]+2\Delta_{0}f(\Omega_p)}, \label{eq:Tp_app}
\end{eqnarray}
Experimentally, the amplitude and phase of the probe transmission are calibrated from the measurement of the amplitude and phase of cavity transmission $T$.

Combining Eq. (\ref{q})-(\ref{A-}), we can also get the solution of $q$ and $(A^+)^*$ as follows,
\begin{eqnarray}
q & = & \frac{[-i(\Delta_{0}+\Omega_{\rm p})+\kappa/2]A^{-}-\varepsilon_{\rm p}}{iGa_{0}} \\
(A^{+})^{*} & = & \frac{-iGa_{0}^{*}q}{i(\Delta_{0}-\Omega_{\rm p})+\kappa/2} \label{APS}
\end{eqnarray}
According to Eq. (\ref{A-}) and Eq. (\ref{APS}), we have the following relation between $A^-$ and $(A^{+})^{*}$,
\begin{eqnarray}
(A^{+})^{*} & = & \eta A^-
\end{eqnarray}
where
\begin{eqnarray}
\eta & = & (\eta_2-\eta_1)\eta_3\\
\eta_1  & = & \frac{[-i(\Delta_{0}+\Omega_{\rm p})+\kappa/2]+2\Delta_{0}f(\Omega_{\rm p})}{1+if(\Omega_{\rm p})} \\
\eta_2 & = & -i(\Delta_{0}+\Omega_{\rm p})+\kappa/2 \\
\eta_3 & = & \frac{-iGa_{0}^{*}}{i(\Delta_{0}-\Omega_{\rm p})+\kappa/2}
\end{eqnarray}
We can therefore calibrate the amplitude ($t_p^{amp}$) and phase ($t_p^{phase}$) of probe transmission based upon the measured cavity transmission $T$ using the following equations,
\begin{eqnarray}
t_p^{amp} &=& \left\vert \frac{a_0}{a_{0}^{*}+a_{0}\eta} \right\vert T^{amp}\\
t_p^{phase} &=& \rm arg \frac{a_0}{a_{0}^{*}+a_{0}\eta} + \phi_T
\end{eqnarray}
where the $T^{amp}$ is the amplitude of $T$.

\section{A simplified solution}
If we assume $(A^{+})^{*}\approx0$, we have a simplified version of the solution. Though this approximation might not be valid in our case, the solution can still help us understand how the photothermal effects influence the system. Equations (\ref{q})-(\ref{AP}) are reduced to
\begin{eqnarray}
(i\Omega_{\rm p}-\gamma_{\rm th})q & = & \gamma_{\rm th}\alpha a_{0}^{*}A^{-}\\
(-i(\Delta_{0}+\Omega_{\rm p})+\kappa/2)A^{-} & = & iGa_{0}q+\varepsilon_{\rm p}
\end{eqnarray}
We obtain the solution as
\begin{eqnarray}
A^{-} & = & \frac{\varepsilon_{\rm p}}{(-i(\Delta_{0}+\Omega_{\rm p})+\kappa/2)-\frac{iG\varepsilon\alpha\left|a_{0}\right|^{2}}{(i\Omega_{\rm p}-\gamma_{\rm th})}} \nonumber\\
 & = & \frac{\varepsilon_{\rm p}}{-i(\Delta_{0}+\Omega_{\rm p}-\frac{\gamma_{\rm th}^{2}G\alpha\left|a_{0}\right|^{2}}{\Omega_{\rm p}^{2}+\gamma_{\rm th}^{2}})+(\kappa/2-\frac{G\varepsilon\alpha\left|a_{0}\right|^{2}\Omega_{\rm p}}{\Omega_{\rm p}^{2}+\gamma_{\rm th}^{2}})}
\end{eqnarray}
From the equation above, we can find the cavity detuning is shifited
by $-\frac{\gamma_{\rm th}^{2}G\alpha\left|a_{0}\right|^{2}}{\Omega_{\rm p}^{2}+\gamma_{\rm th}^{2}}$
and the cavity decay is modified by the photothermal effects by
\begin{eqnarray}
\kappa_{th} & = & -\frac{G\varepsilon\alpha\left|a_{0}\right|^{2}\Omega_{\rm p}}{\Omega_{\rm p}^{2}+\gamma_{\rm th}^{2}}
\end{eqnarray}
When $\Omega_{\rm p}$ approaches zeros, we have a window of Lorentzian shape which can be described by the normalized reflectivity as follows.
 \begin{eqnarray}
A^{-} & = & \frac{\varepsilon_{\rm p}(i\Omega_{\rm p}-\gamma_{\rm th})}{\kappa(i\Omega_{\rm p}-\gamma_{\rm th})/2-iG\gamma_{\rm th}\alpha\left|a_{0}\right|^{2}} \\
r & = & 1-\frac{k(i\Omega_{\rm p}-\gamma_{\rm th})/2}{\kappa(i\Omega_{\rm p}-\gamma_{\rm th})/2-iG\gamma_{\rm th}\alpha\left|a_{0}\right|^{2}} \nonumber\\
 & = & \frac{-iG\gamma_{\rm th}\alpha\left|a_{0}\right|^{2}}{\kappa(i\Omega_{\rm p}-\gamma_{\rm th})/2-iG\gamma_{\rm th}\alpha\left|a_{0}\right|^{2}}\\
|r|^{2} & = & \frac{4G^{2}\gamma_{\rm th}^{2}\alpha^{2}\left|a_{0}\right|^{4}/\kappa^{2}}{(2G\gamma_{\rm th}\alpha\left|a_{0}\right|^{2}/\kappa-\Omega_{\rm p})^{2}+\gamma_{\rm th}^{2}}\nonumber\\
 & = & \frac{C^{2}}{(C-\Omega_{\rm p}/\gamma_{\rm th})^{2}+1}\\
C & = & 2G\alpha\left|a_{0}\right|^{2}/\kappa
\end{eqnarray}
The reflectivity exceeds one in the case of $C>1$
which can be easily reailzed by enhancing the pump power. The Lorentzian
of width is
\begin{eqnarray}
\Gamma_{th} & = & 2\gamma_{\rm th}
\end{eqnarray}
The reflection amplitude and phase under this limit take the form
\begin{eqnarray}
r  & = & \frac{-C(\Omega_{\rm p}/\gamma_{\rm th}-C)+iC}{(\Omega_{\rm p}/\gamma_{\rm th}-C)^{2}+1} \\
\phi_{r}(\Omega_{\rm p}) & = & \rm arctan(-\frac{1}{\Omega_{\rm p}/\gamma_{\rm th}-C})
\end{eqnarray}

The reflection group delay is given as
 \begin{eqnarray}
\tau_{r} & = & -\frac{d\phi_{r}(\Omega_{\rm p})}{d\Omega_{\rm p}} \nonumber\\
 & = & \frac{1}{1+(\frac{1}{\Omega_{\rm p}/\gamma_{\rm th}-C})^{2}}\frac{1/\gamma_{\rm th}}{(\Omega_{\rm p}/\gamma_{\rm th}-C)^{2}} \nonumber\\
 & = & \frac{1/\gamma_{\rm th}}{(\Omega_{\rm p}/\gamma_{\rm th}-C)^{2}+1}\\
\tau_{r} & = & \mathbf{R}\{\frac{-i}{r}\frac{dr}{d\Omega_{\rm p}}\}
\end{eqnarray}

The transmission group delay is obtained as
\begin{eqnarray}
t & = & 1-\frac{-iC}{i(\Omega_{\rm p}/\gamma_{\rm th}-C)-1}\\
\tau_{r} & = & \mathbf{R}\{\frac{-i}{t}\frac{dt}{d\Omega_{\rm p}}\} \nonumber\\
 & = & \mathbf{R}\{\frac{C\varepsilon(\gamma_{\rm th}+i\Omega_{\rm p})(i\varepsilon-\Omega_{\rm p}+C\varepsilon)}{(\gamma_{\rm th}^{2}+\Omega_{\rm p}^{2})(C^{2}\gamma_{\rm th}^{2}-2C\varepsilon\Omega_{\rm p}+\gamma_{\rm th}^{2}+\Omega_{\rm p}^{2})}\} \nonumber\\
 & = & \mathbf{R}\{\frac{1}{\gamma_{\rm th}}\frac{C(1+i\Omega_{\rm p}/\gamma_{\rm th})(i-\Omega_{\rm p}/\gamma_{\rm th}+C)}{[1+(\Omega_{\rm p}/\gamma_{\rm th})^{2}][(\Omega_{\rm p}/\gamma_{\rm th}-C)^{2}+1\text{]}}\} \nonumber\\
 & = & \mathbf{R}\{\frac{1}{\gamma_{\rm th}}\frac{C(-2\Omega_{\rm p}/\gamma_{\rm th}+C)}{[1+(\Omega_{\rm p}/\gamma_{\rm th})^{2}][(\Omega_{\rm p}/\gamma_{\rm th}-C)^{2}+1\text{]}}\}
\end{eqnarray}
\end{document}